\documentclass{moriond}
\def\Journal#1#2#3#4{{#1} {\bf #2}, #3 (#4)}

\def\PRL{\em Phys. Rev. Lett.}
\def\PRD{{\em Phys. Rev.} D}

\def\JHEP{\em JHEP}
\def\EPJC{{\em Eur. Phys. J.} C}

\newcommand {\afb} {A_\mathrm{FB}}
\newcommand {\sineff} {\sin^2\theta_\mathrm{eff}^\ell}
\newcommand{\zll} {Z/\gamma\rightarrow\ell\ell}
\newcommand{\cuu} {$\mu\mu$}
\newcommand{\cee} {$ee$}
\newcommand{\ceg} {$eg$}
\newcommand{\ceh} {$eh$}
\newcommand{\cs}{$\cos\theta_\mathrm{CS}$ }
\newcommand{\mycite}{\,\cite}
\newcommand{\twofigs}{0.375\linewidth}

\newcommand{\Photo}{}

\begin{document}
\vspace*{4cm}
\title{Measurement of the effective leptonic electroweak mixing angle and Drell--Yan forward-backward asymmetry using pp collisions at $\sqrt{s}=13$ TeV }

\author{ Aleko Khukhunaishvili  \\ on behalf of the CMS Collaboration}
\address{University of Rochester}

\maketitle
\abstracts{
A measurement of the effective leptonic weak mixing angle $\sineff$ is presented using the 
forward-backward asymmetry in Drell--Yan dilepton events produced in pp
collisions at $\sqrt{s}=13$\,TeV. The data sample corresponds to 137\,fb$^{-1}$ of integrated luminosity 
and consists of dimuon and dielectron events, including the forward electrons reconstructed beyond the coverage 
of the CMS tracking detectors. Using the CT18Z set of the parton distribution functions\,(PDF), which provides 
the best coverage of the central values extracted with the other global PDFs, we obtain
$$\sin^2\theta^\ell_\mathrm{eff} = 0.23157 \pm 0.00031,$$
where the total uncertainty includes the statistical uncetainties\,(0.00010), as well as the correlated
experimental\,(0.00015), theoretical\,(0.00009), and PDF\,(0.00027) systematic uncertainties. 
The measured value agrees well with the Standard Model prediction. 
This is the most precise measurement at a hadron collider with comparable precision to the 
two most precise results obtained at LEP and SLD. 
}

\section{Introduction}
The electroweak mixing angle, $\theta_\mathrm{W}$, is a key Standard Model\,(SM) parameter 
that was introduced to explain the masses of the W and Z bosons through the spontaneous 
electroweak symmetry breaking. It thereby relates their masses,  
$\sin^2\theta_\mathrm{W} = 1-m^2_\mathrm{W}/m^2_\mathrm{Z}$, and defines the ratio between the vector and 
axial-vector couplings of the Z bosons with the fermions. Beyond the tree level, \textit{effective} weak mixing angles 
are defined based on these couplings and include the fermion-flavor dependent radiative corrections.  

We measure\mycite{thismeas} the \textit{leptonic} effective weak mixing angle, $\sineff$, using the Drell--Yan dilepton 
events produced in pp collisions at 13 TeV. The two most precise previous measurements of this 
parameter, performed by the LEP and SLD collaborations\mycite{lepsld} in electron-positron collisions,  differ by 
about 3 standard deviations. More recently, the measurements at the hadron colliders have also become
competitive\mycite{lhcbsw2,cdfsw2,d0sw2,cmssw2,atlassw2}.

Precise measurement of $\sineff$ provides a crucial test of the SM since the theory accurately 
predicts its value of $0.23155\pm0.00004$\mycite{pdg} using other precise experimental inputs.
Therefore, a significant deviation from this prediction would indicate that the SM relations need to be adjusted to include 
new physics contributions. Moreover, $\sineff$ is closely related to $m_\mathrm{W}$, for which the CDF collaboration 
recently reported a very precise measurement\mycite{cdfmw}, which disagrees with both the previous
measurements and the SM. The relation between these two parameters is such that models\mycite{2hdm} that describe 
the large CDF value of $m_\mathrm{W}$  would predict a lower $\sineff$ value than the SM. 

\section{Method description and measurement details}
The $\sineff$ is measured using $\zll$ events, where $\ell$  is either muon or electron, in a wide dilepton mass window 
of 54--150~GeV around the Z boson mass. In most of these events, a quark from one proton interacts with its antiquark in 
the other, producing a Z boson, which decays into a lepton pair:
$q\bar{q}\rightarrow \mathrm{Z}/\gamma\rightarrow\ell^-\ell^+$.
In these events, the angle $\theta$ between the $\ell^-$ and $q$ directions is distributed as  
$1+\cos^2\theta + 0.5 A_0 (1-3\cos^2\theta) + A_4\cos\theta$, where
the last term defines the asymmetry $\afb=3/8A_4$ between the forward~($\cos\theta>0$) and backward~($\cos\theta<0$) 
cross sections. For the dilepton mass values near $m_\mathrm{Z}$, this asymmetry strongly depends on  $\sineff$\mycite{cmssw2}. 

To reduce theoretical and experimental uncertainties, the lepton decay angles are computed in the Colins-Sopper 
center of mass frame. In pp collisions, the positive z-axis direction of this frame is based on the longitudinal boost of the 
dilepton system,  relying on the fact that on average it corresponds to the direction of the valence quark. 
Two consequences follow: (1) only the valence-quark production channels contribute to the observed $\afb$, which (2) 
has a significant dependence on the dilepton rapidity resulting from the wrong guess of the quark direction. 
As a result, the observed $\afb$ also has a strong dependence on PDFs, especially at high and low dilepton masses.
The PDF uncertainties, which dominate the uncertainties of $\sineff$ measurements in pp collisions, are reduced
by measuring the $\afb$ as a function of the dilepton mass and rapidity and thereby constraining the PDFs in situ\mycite{pdfinsitu}.

The default method, which was also used in Run 1\mycite{cmssw2}, is based on the observed angular-weighted $\afb$\mycite{afbw}, 
which is less sensitive to the systematic uncertainties in the event selection efficiencies. We also, however, extract the $\sineff$ from the 
intermediate  $A_4$ measurements unfolded in the pre-FSR dilepton rapidity~(Y) and mass~(M) bins. These $A_4(Y, M)$ results can 
be used for future re-extraction of $\sineff$, e.g. with improved PDFs, and in combination with other electroweak measurements with 
common theoretical and PDF uncertainties. 

\begin{figure}[!htb]
\centering
\includegraphics[width=0.325\linewidth]{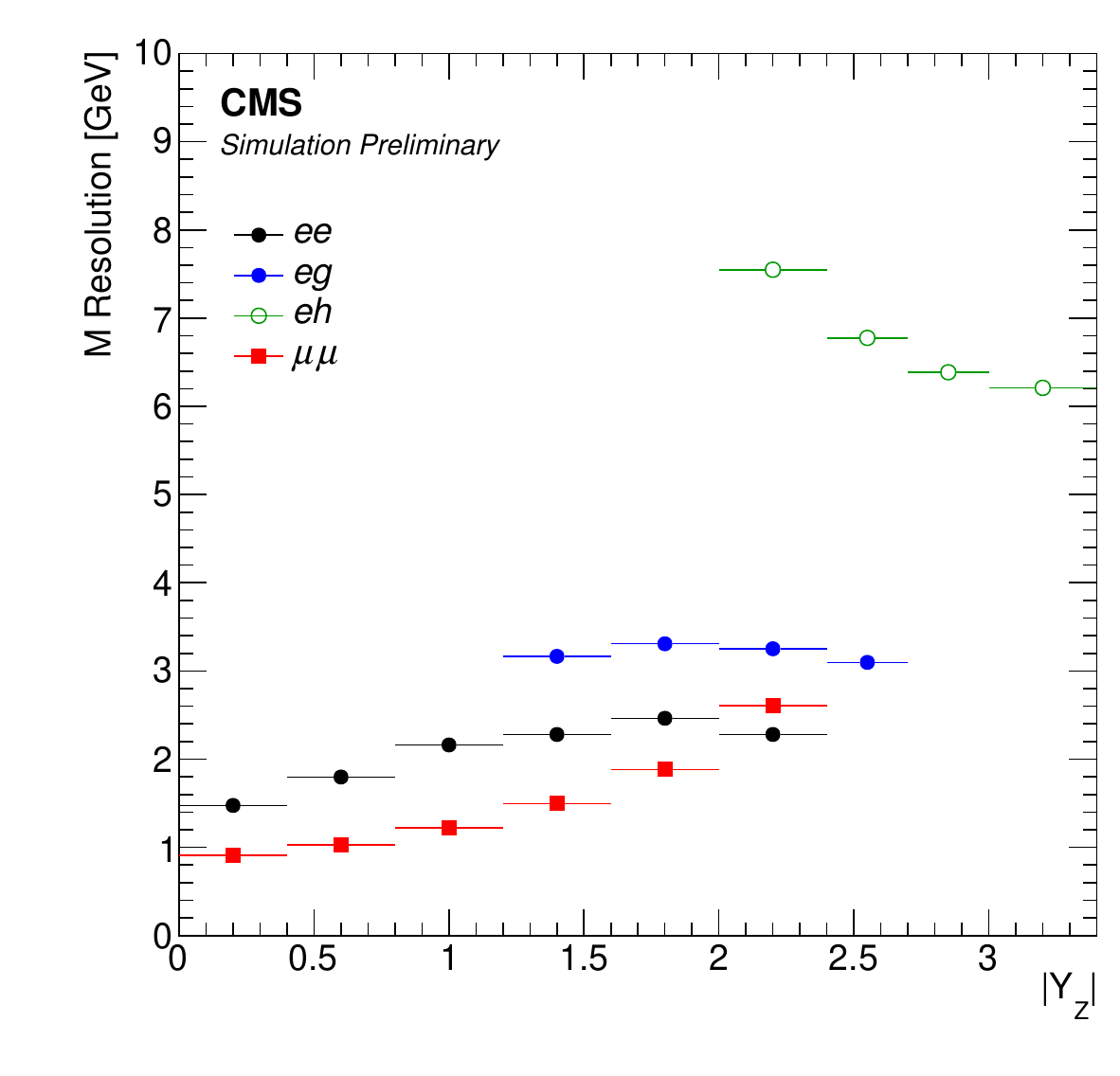}
\includegraphics[width=0.325\linewidth]{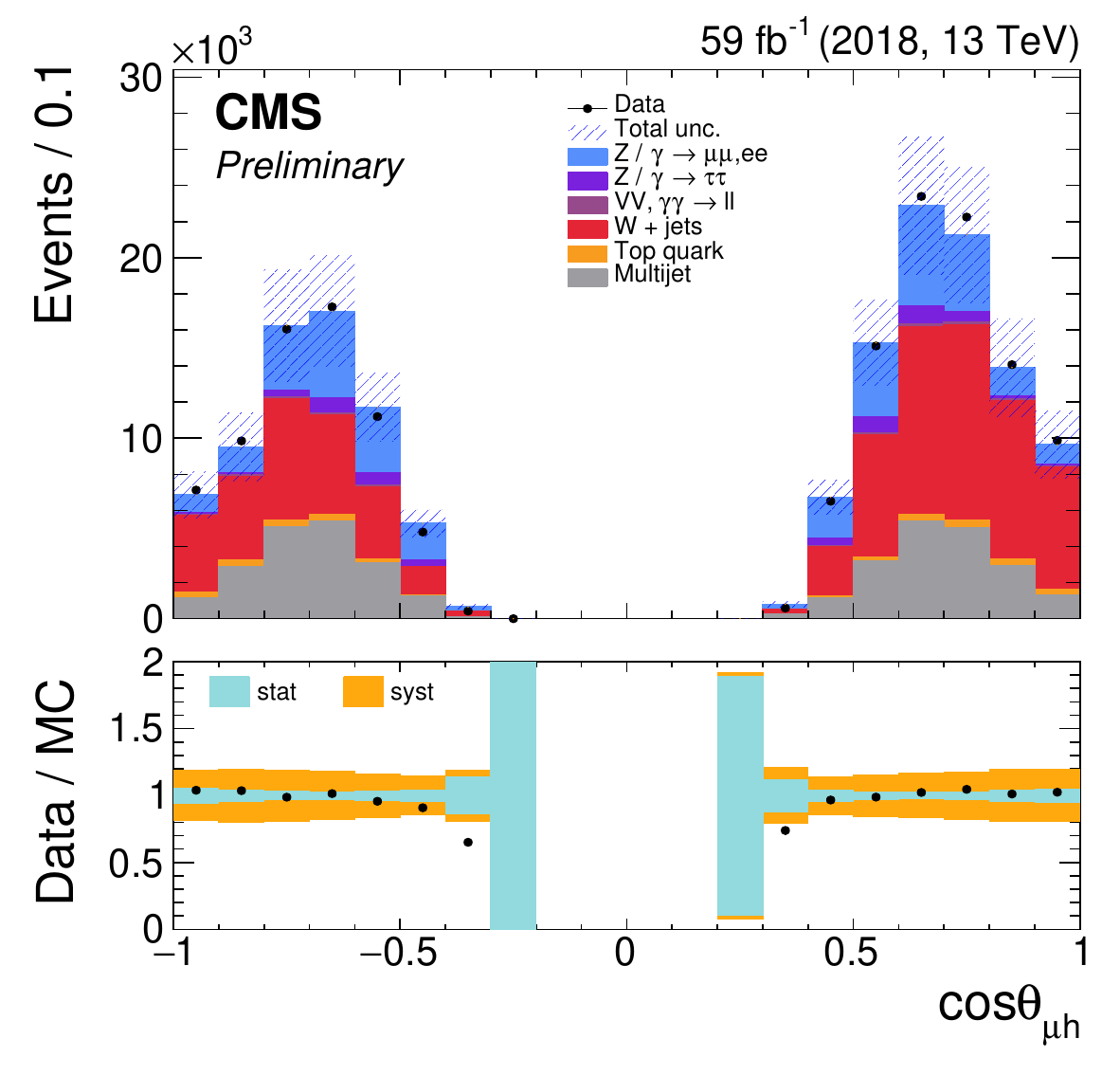}
\includegraphics[width=0.325\linewidth]{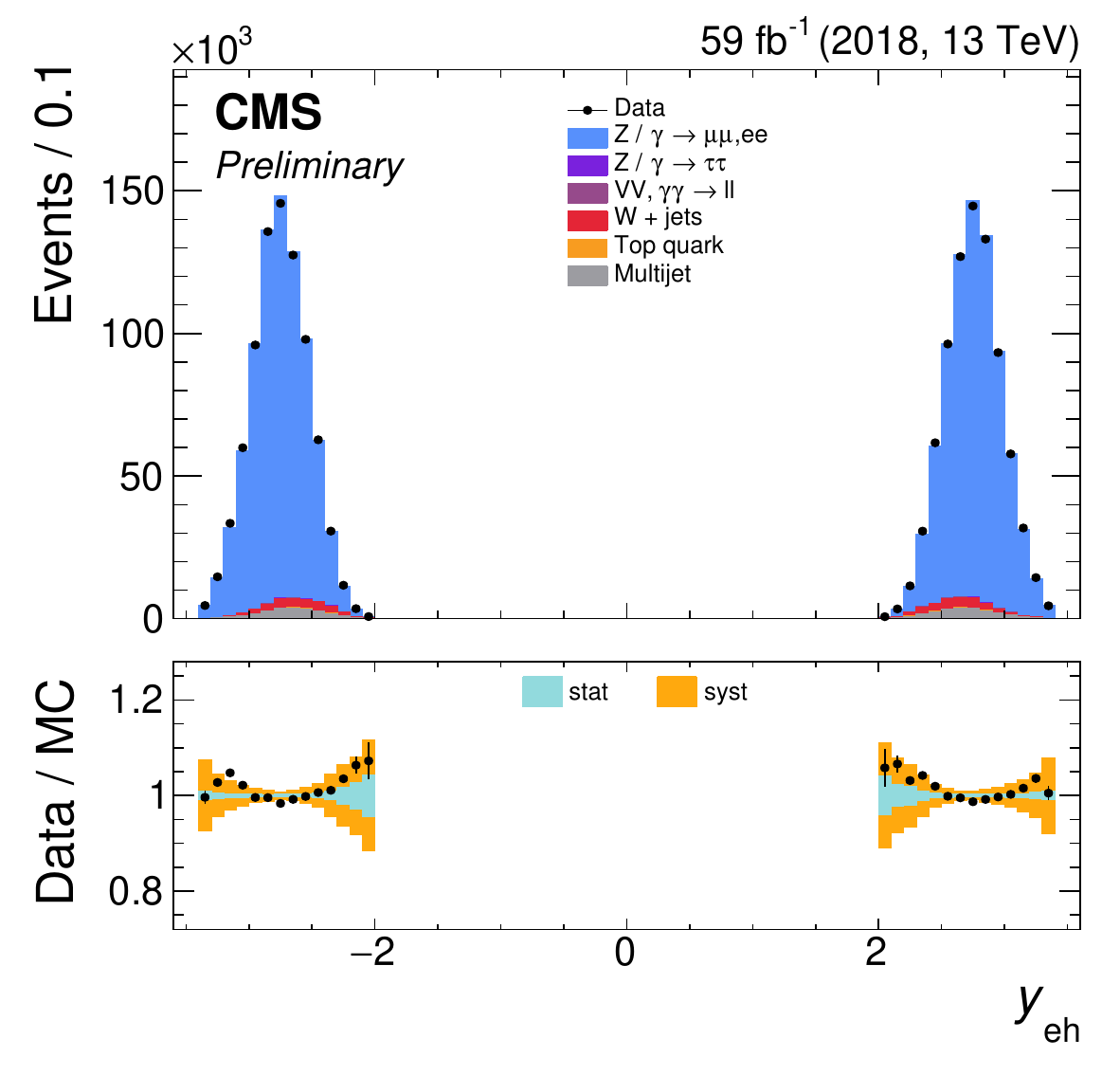}
\caption[]{Dilepton mass resolutions in four channels~(left),  $\cos\theta$ distribution in $\mu h$ control region~(middle), and rapidity distribution in the \ceh ~channel~(right). }
\label{fig:ak1}
\end{figure}

The measurement uses four types of reconstructed dilepton events: \cuu, \cee, \ceg, and \ceh, where 
$\mu$ stands for muons\mycite{muons} reconstructed in the pseudorapidity range $|\eta|<2.4$, 
$e$ stands for the central electrons\mycite{electrons} reconstructed within the tracking detectors: $|\eta|<2.5$,
$g$ stands for the ``EF'' electrons reconstructed in beyond-the-tracker region of the electromagnetic calorimeter: $2.50<|\eta|<2.87$, and
$h$ stands for the ``HF'' electrons reconstructed in the forward hadronic calorimeter: $3.14<|\eta|<4.36$.
For the EF electrons, dedicated cut-based identification criteria were developed in TMVA\mycite{tmva} for this analysis. 
The HF electrons are reconstructed from the electromagnetic and hadronic constituents of the forward jets. 
The multivariate regression and classification techniques, implemented in \textsc{TensorFlow}\mycite{tflow}, were developed to 
optimize their momentum reconstruction and distinguish them from the hadronic jets. 
As shown in Figure~\ref{fig:ak1}~(left), the dilepton mass resolutions range from about 1 GeV for the central-rapidity dimuons 
to about 7 GeV for the \ceh ~dielectrons. 

The multijet background is estimated with transfer factors applied to the multijet-enriched regions. 
For the \cuu ~and \cee ~channels, the method is independently validated using the same-sign dilepton events. 
The W+jets background is estimated from the simulation. However, in the \ceg ~and \ceh ~channels, where its contribution is more important, 
it is corrected using the fake-lepton scale factors, obtained in the $ug$ and $uh$ control regions. The measured and predicted $\cos\theta$ 
distributions in the $uh$ sample, with the corrected multijet and W+jets background normalizations, are shown in Figure~\ref{fig:ak1}
~(middle). Other small electroweak and top quark backgrounds are evaluated from simulation and validated in the $e\mu$ control regions. 

The signal events are simulated with \textsc{Powheg} MiNNLO\mycite{minnlo} program, interfaced with
\textsc{Pythia8} for the parton showering and hadronization,  as well as for the initial-state photon radiation\mycite{pythia}. 
The final-state photon radiation~(FSR) is modeled with \textsc{Photos}\mycite{photos}. To improve the signal model, 
various corrections are applied: next-to-leading order~(NLO) weak corrections, pileup reweighting, trigger-prefiring corrections, 
lepton selection efficiency scale factors, electron charge-misidentification scale factors, lepton-momentum corrections, and dilepton  
$p_\mathrm{T}$ reweighting.

\begin{figure}[!htb]
\centering
\includegraphics[width=\twofigs]{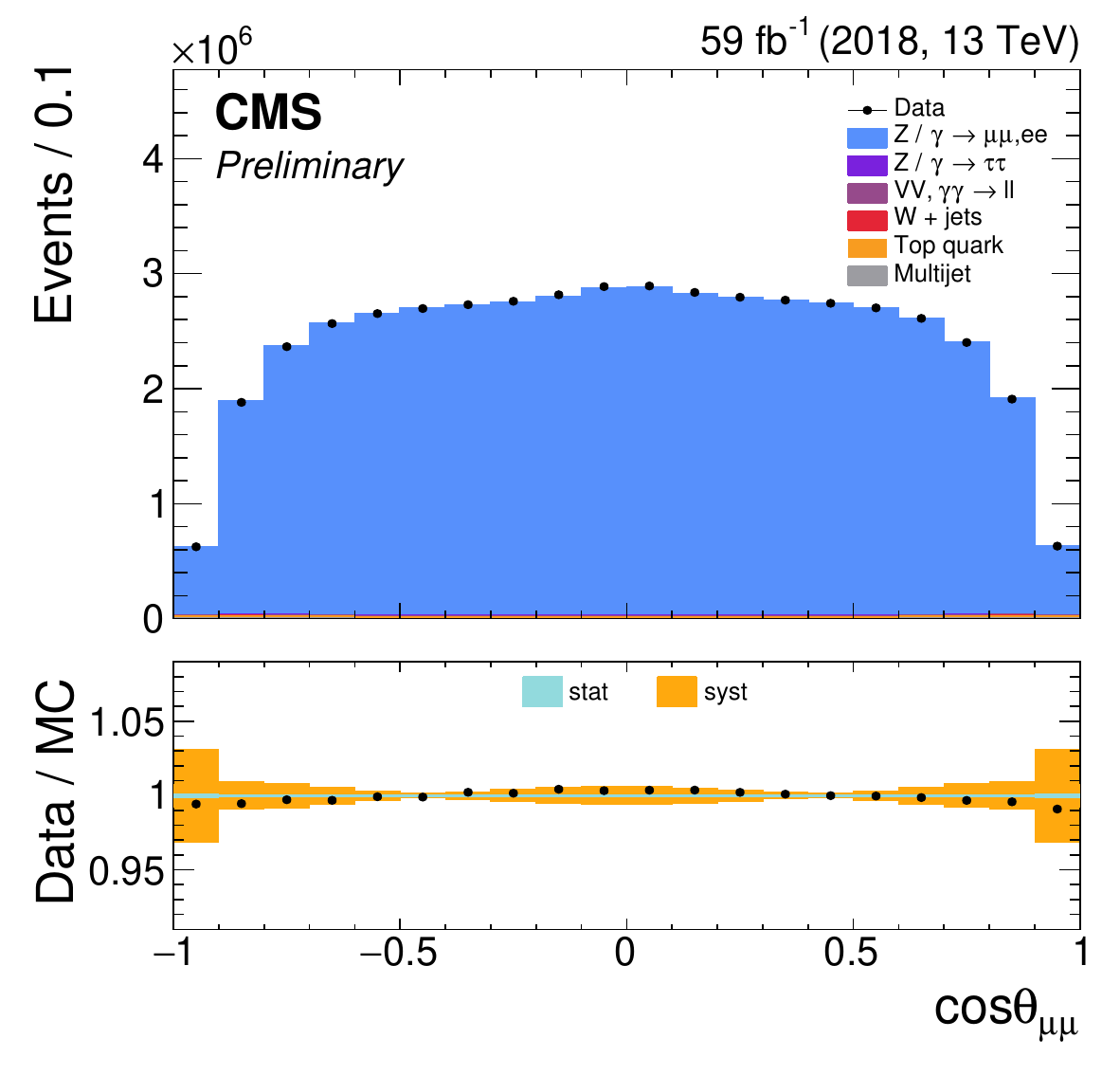}
\includegraphics[width=\twofigs]{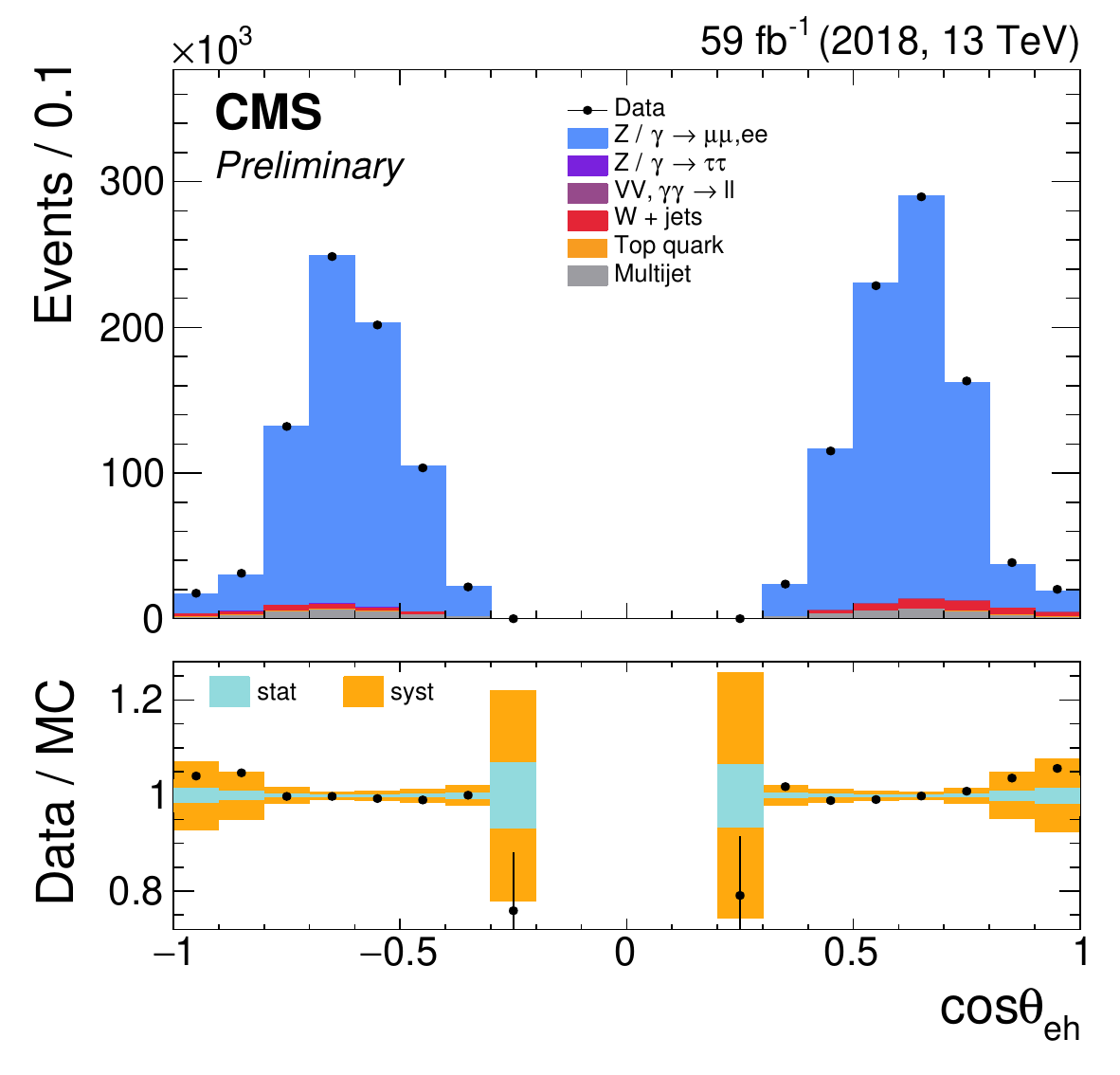}
\caption[]{Lepton $\cos\theta$ distribution in \cuu~(left) and \ceh~(right) dilepton channels.}
\label{fig:ak2}
\end{figure}

Figures~\ref{fig:ak1}~(right)  and \ref{fig:ak2} show excellent agreement between the measured and predicted distributions. 
The default configuration of the signal samples includes the NLO and universal higher-order~(HO) weak
corrections\mycite{nloew} obtained using the \textsc{Powheg-EW} program. The default input scheme\mycite{sw2input} 
uses $\sineff$, $m_\mathrm{Z}$, and the Fermi constant $G_\mu$, while the finite-width effects are implemented according 
to the complex-mass scheme. 

\begin{figure}[!htb]
\centering
\begin{minipage}{\twofigs}
\includegraphics[width=\linewidth,height=0.9\linewidth]{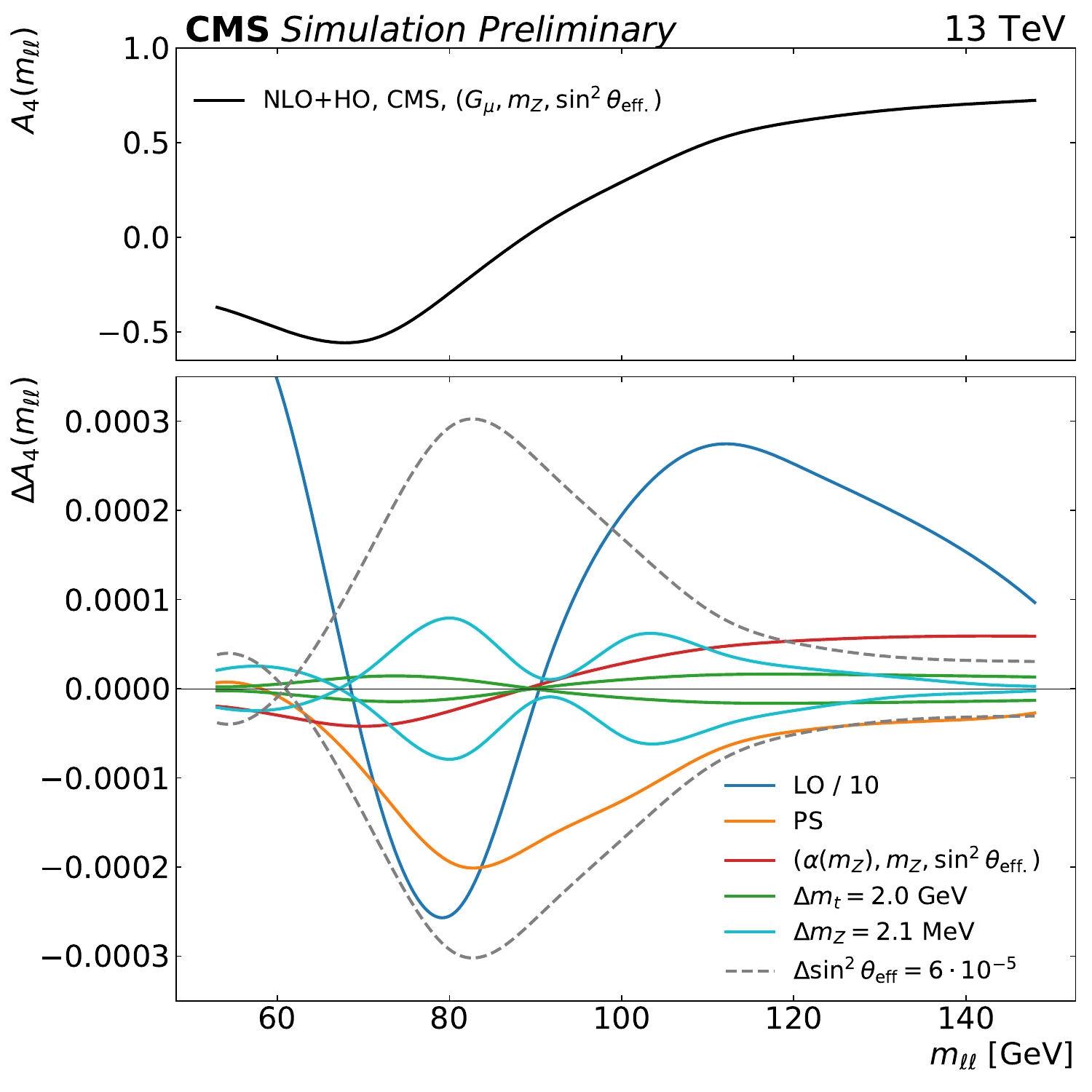}
\end{minipage}
\begin{minipage}{\twofigs}
\includegraphics[width=\linewidth]{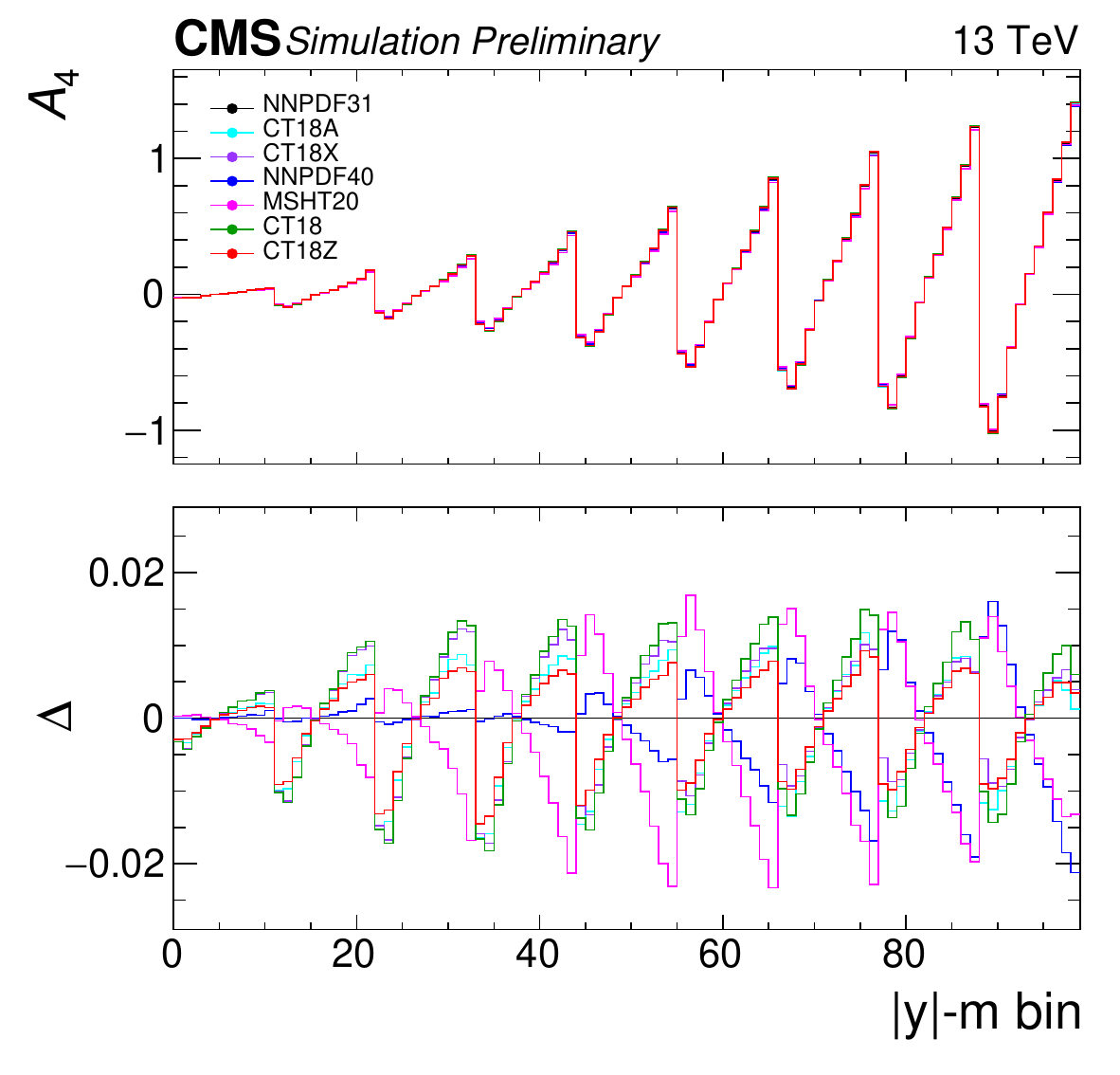}
\end{minipage}
\caption[]{Dependence of the $A_4$ angular coefficient on the electroweak configuration settings~(left) and PDFs~(right).}
\label{fig:ak3}
\end{figure}

Theoretical systematic sources include renormalization and factorization scales, dilepton $p_\mathrm{T}$ modeling,  
FSR,  alternative propagator width and electroweak input schemes, as well as the corresponding input parameter values. 
We also consider several modern PDF sets\mycite{NNPDF31,NNPDF40,CT18,MSTH20} and corresponding uncertainties. 
The dependence of the $A_4$ coefficient on the electroweak input settings and PDFs is shown in Figure~\ref{fig:ak3}.
Experimental systematic uncertainties include the MC statistical uncertainties, as well as the uncertainties in the lepton selection
efficiencies, momentum calibration, backgrounds, and other sources~(trigger prefiring, vertex distribution,  charge misidentification, pileup, and integrated luminosity). 

\section{Fits and results}
\begin{figure}[!htb]
\centering
\includegraphics[width=\twofigs]{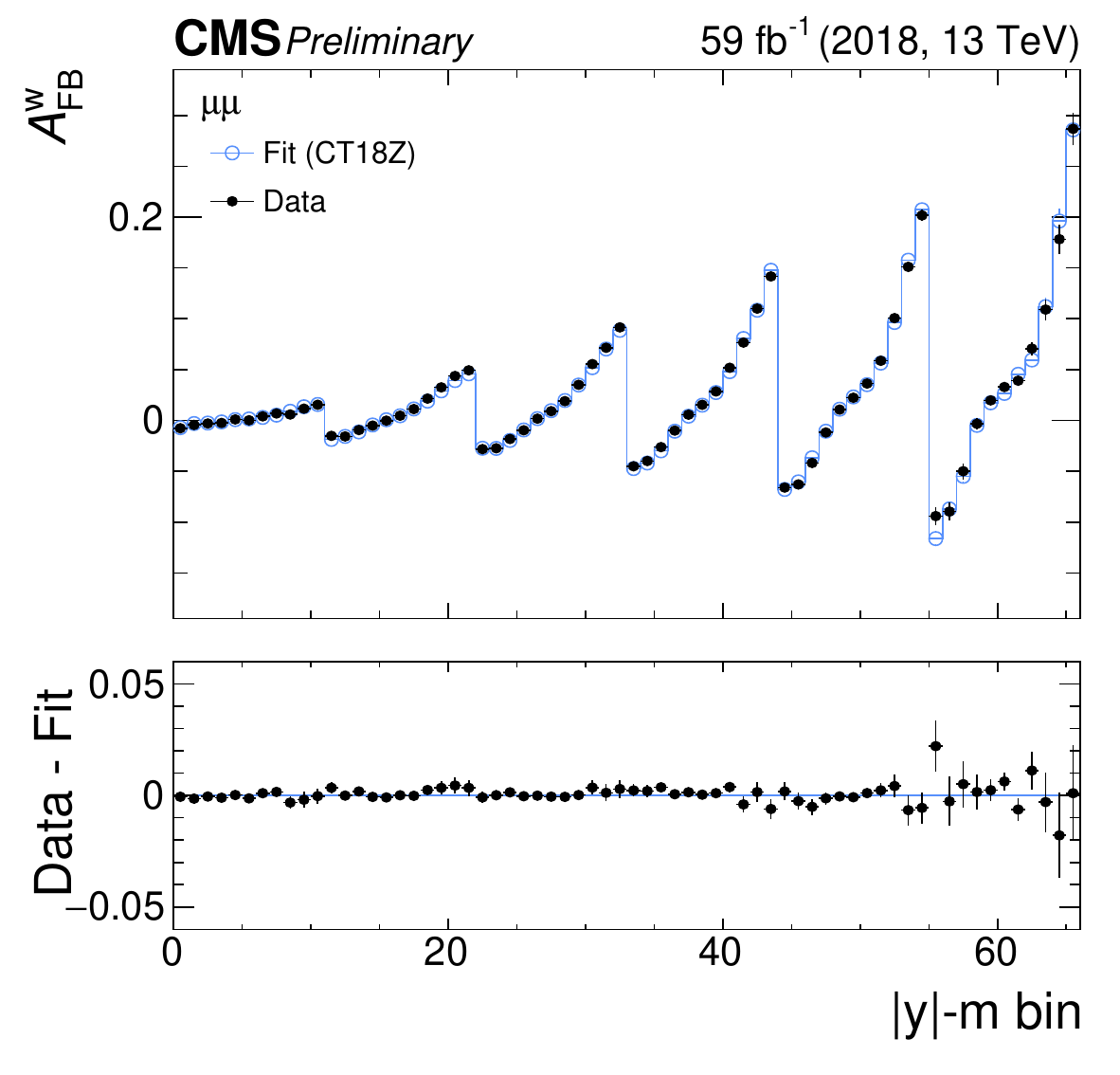}
\includegraphics[width=\twofigs]{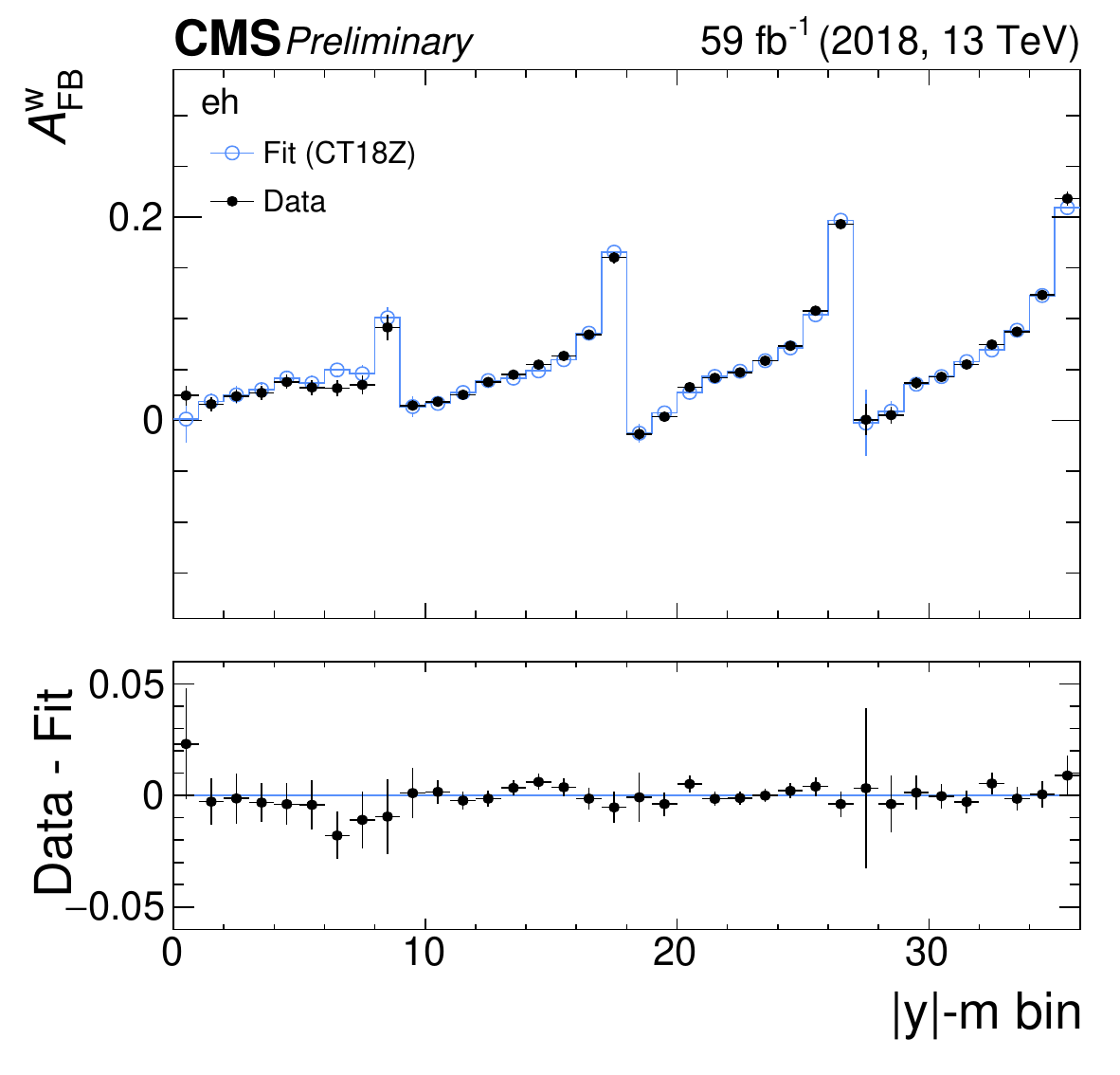}
\caption{Measured and best-fit angular weighted $\afb(y,m)$ distributions 
for the 2018 period and in the \cuu~(left) and \ceh~(right) channels.
The error bars represent the statistical uncertainties of the measured and simulated samples.}
\label{fig:ak4}
\end{figure}

The $\sineff$ is extracted by a simultaneous $\chi^2$-fit of the observed $\afb(y,m)$ in all years and channels with the 
corresponding templates. The fit results for the four channels and their combination are summarized 
in Table~\ref{fig:ak4}. The comparison between the measured and the corresponding best-fit $\afb(y,m)$ distributions for \cuu ~and \ceh ~channels in 2018 is also shown in Figure~\ref{fig:ak4}.

\begin{table}[!htb]
\footnotesize
\centering
\caption{Fit results (in units of $10^{-5}$) for the four channels and for all~($\ell\ell$) channels, 
using the full Run~2 event sample. The experimental systematic uncertainties (``exp") 
include the uncertainties listed in the last five columns.}
\label{tab:ak1}
\begin{tabular}{l c c c c | c c c c | c c c c c}
&  $\chi^2$ & bins & $p$(\%) & $\sineff$  & stat & exp & theo & PDF & MC & bkg & eff & calib & other \\
\hline
\cuu  &  241.3 & 264  &   82.7 &  $23146 \pm 38$ &  17 &  17 & 7 & 30 &  13 &  3 &   2 &   5 &   4 \\
\cee &  256.7 & 264  &   59.8 &  $23176 \pm 41$ &  22 &  18 & 7 & 30 &  14 &  4 &   5 &   3 &   7 \\
\ceg &  119.1 & 144  &   92.8 &  $23257 \pm 61$ &  30 &  40 & 5 & 44 &  23 & 11 & 12 & 19 &   9 \\
\ceh &  104.6 & 144  &   99.3 &  $23119 \pm 48$ &  18 &  33 & 9 & 37 &  14 & 10 & 16 & 18 &   6 \\
$\ell\ell$   &  730.7 & 816  &   98.4 &  $23157 \pm 31$ &  10 &  15 & 9 & 27 &    8 &   4 &   6 &   6 &   3 \\
\hline
\end{tabular}
\end{table}

\begin{figure}[!htb]
\centering
\includegraphics[width=\twofigs]{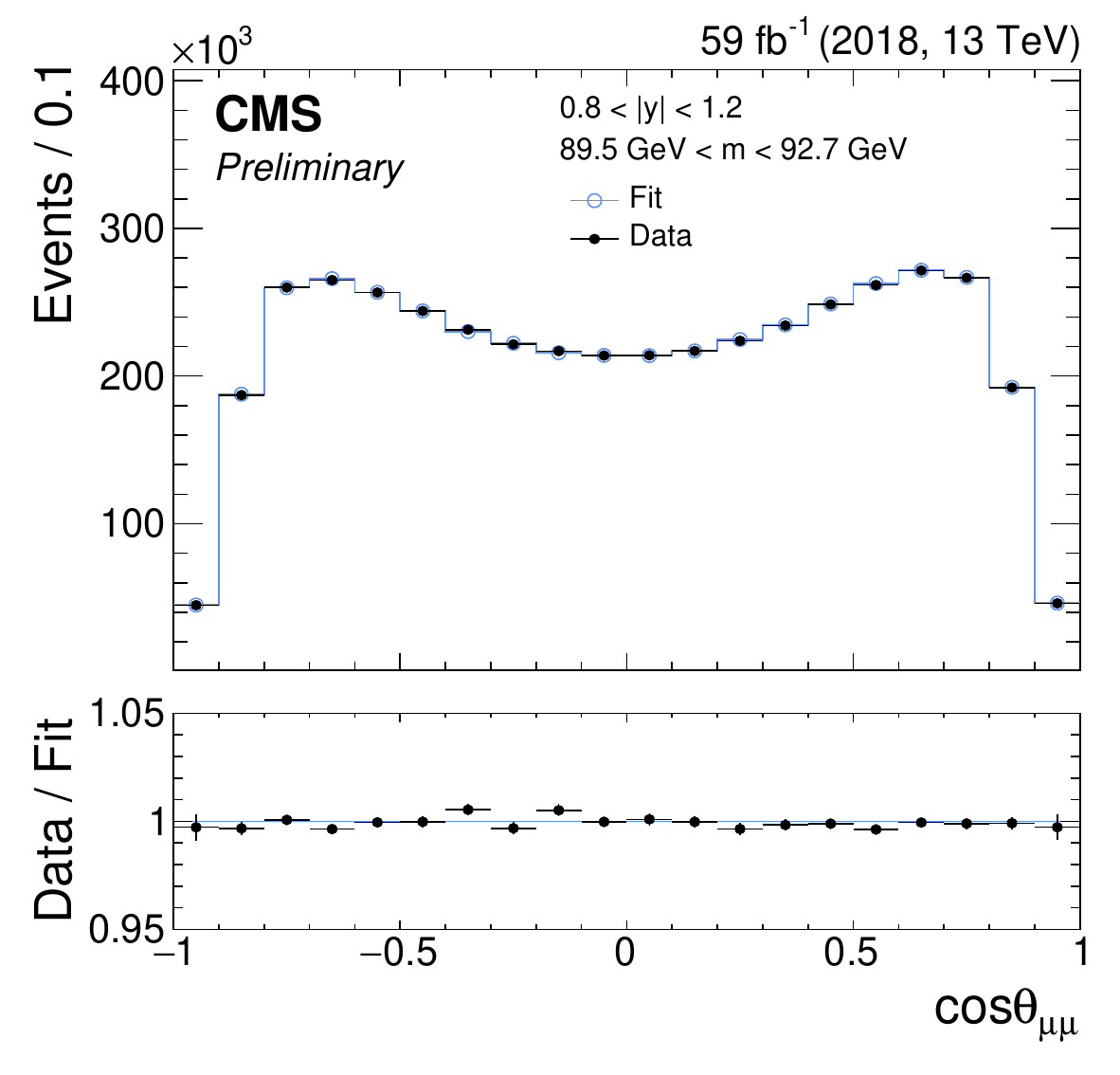}
\includegraphics[width=\twofigs]{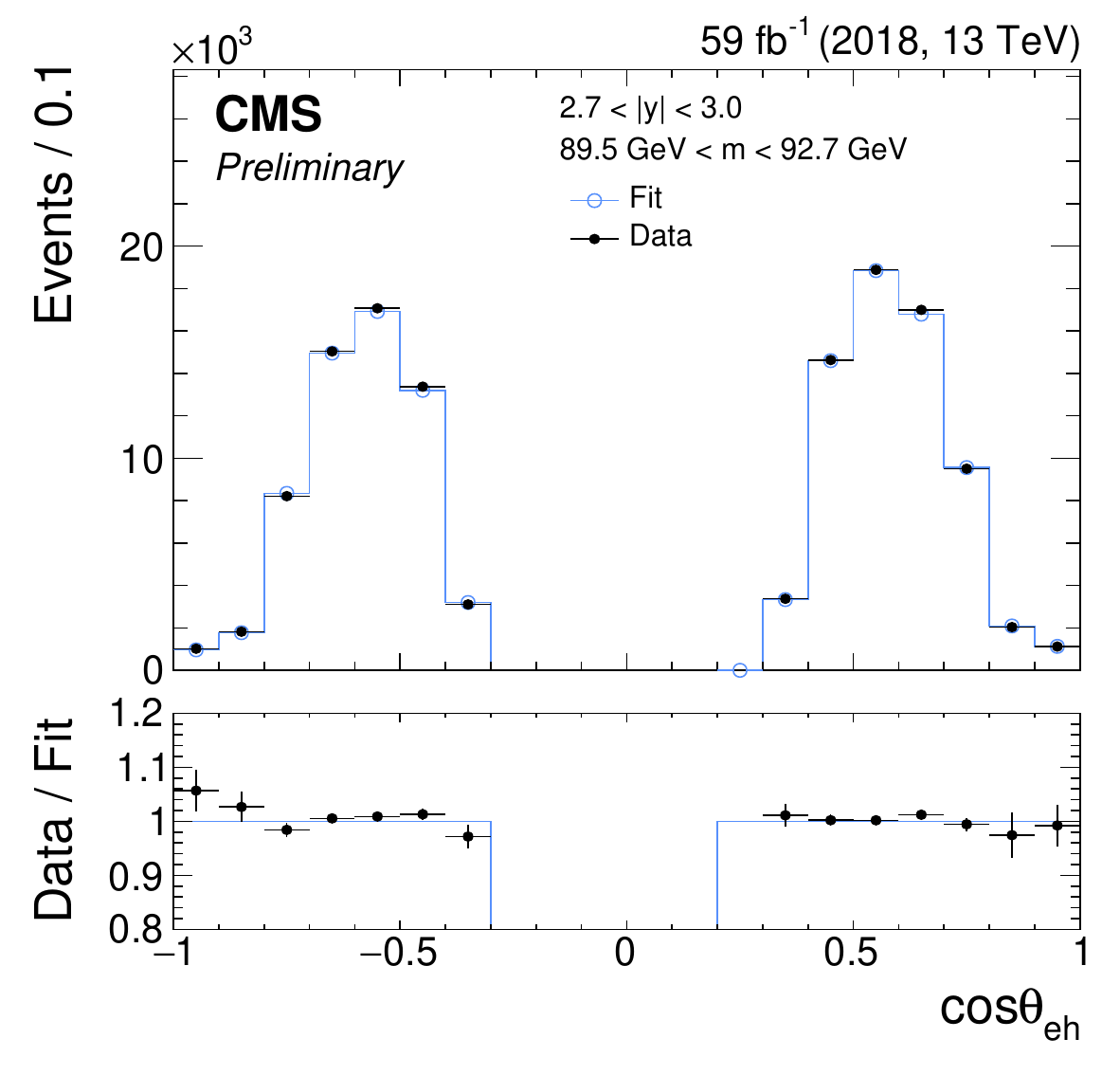}
\caption{Measured and best-fit \cs distributions in the \cuu (left) and \ceh (right) channels of the 2018 samples, 
for the dilepton mass peak and relevant rapidity bins for each channel. 
The error bars represent the statistical uncertainties.}
\label{fig:ak5}
\end{figure}

The $A_4$ angular coefficient is unfolded to the pre-FSR $Y$-$M$ bins by fitting the observed \cs distributions
in the reconstructed $y$-$m$ bins simultaneously in all runs and channels. 
In the combined fit configuration, we have 14205 bins with predicted entries greater than 10, 
101 free parameters of interest, and 3361 nuisance parameters for all systematic uncertainties. 
The minimization is performed with the L-BFGS method\mycite{lbfgs} that relies on the analytic gradient calculations. 
The resulting covariance matrix is also evaluated fully analytically by computing the inverse of 
the Hessian matrix at the minimum.  The \cs distributions in two important example bins, displayed in 
Figure~\ref{fig:ak5}, show an excellent agreement between the data and best-fit distributions. 
The overall $\chi^2_\mathrm{min}=14839$ also indicates a reasonable overall fit quality even 
for the full distributions\,(i.e. not just asymmetries, which are more robust and easier to model).

\begin{figure}[!htb]
\centering
\begin{minipage}{\twofigs}
\includegraphics[width=\linewidth]{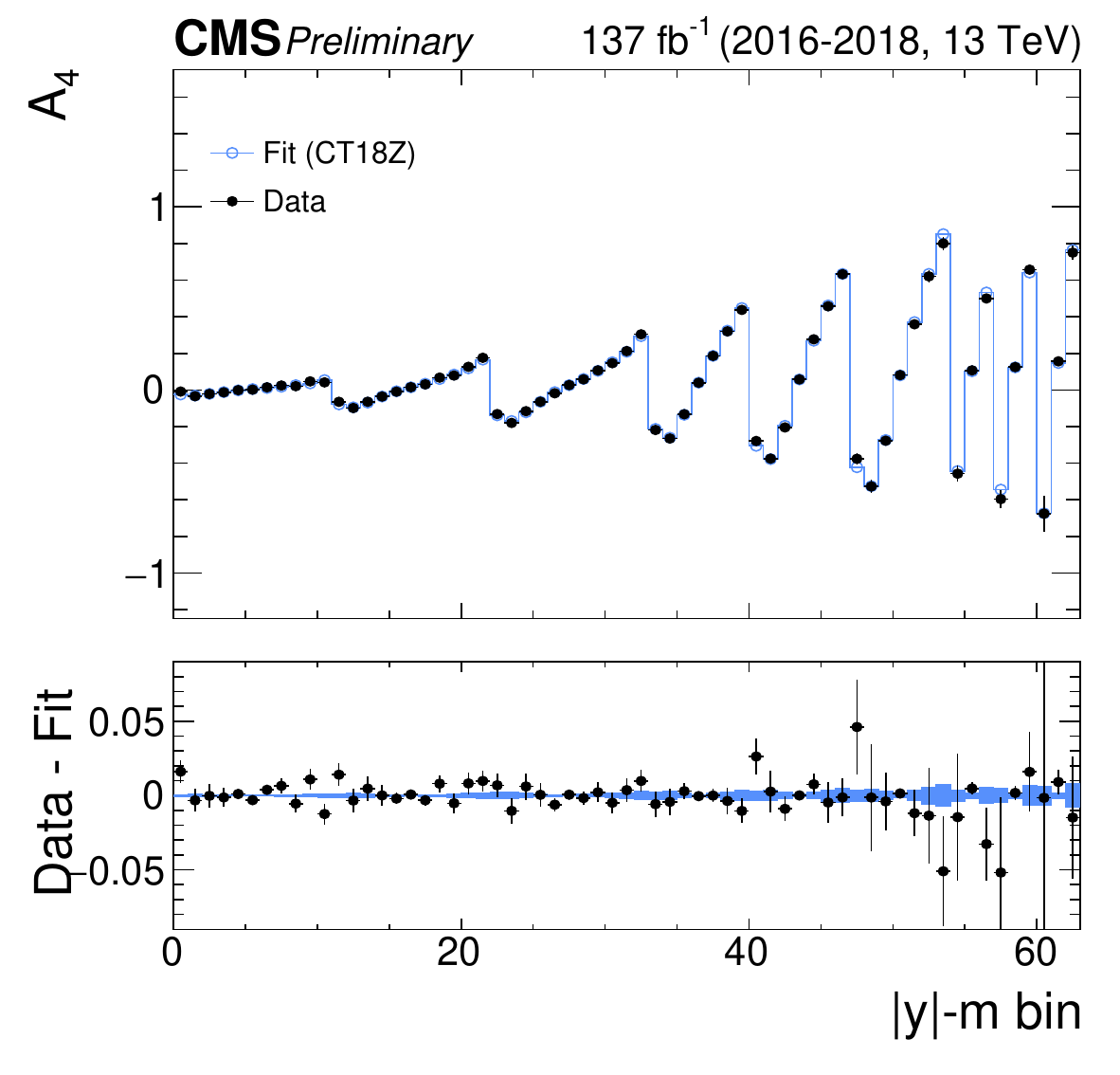}
\end{minipage}
\begin{minipage}{0.57\linewidth}
\includegraphics[width=\linewidth]{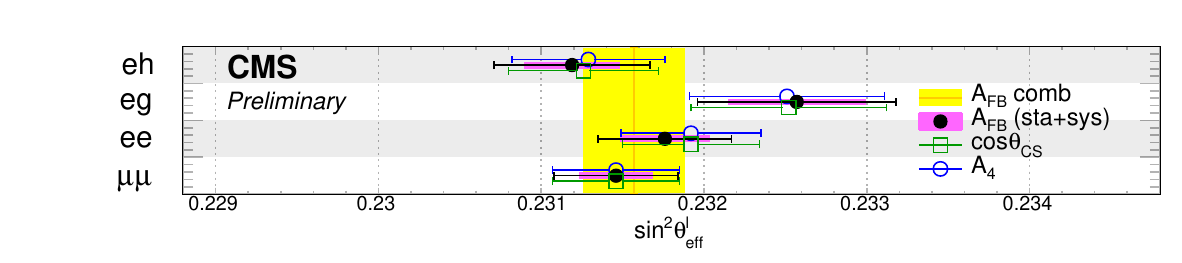}
\includegraphics[width=\linewidth]{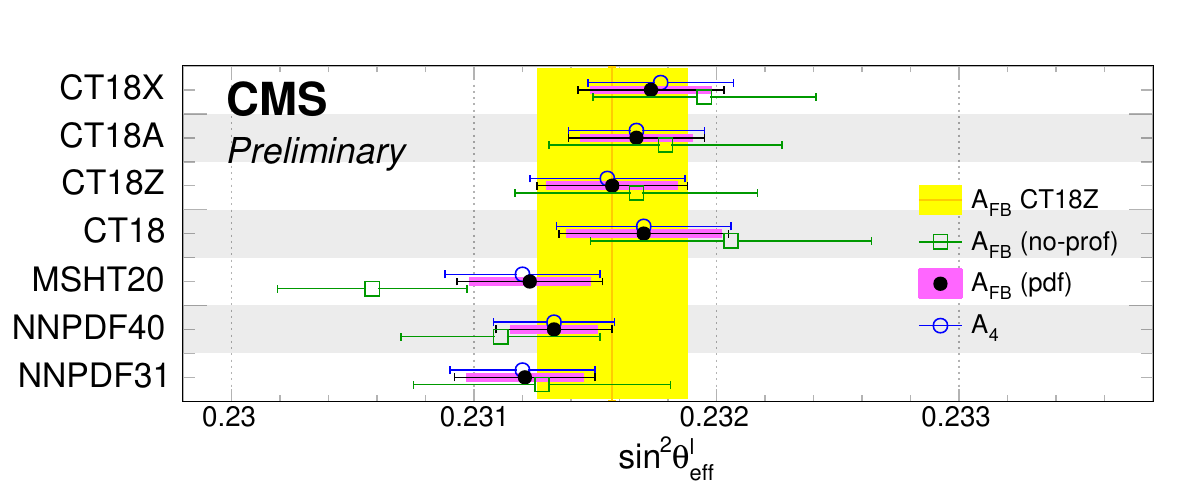}
\end{minipage}
\caption{Left: Measured and best-fit $A_4(Y,M)$ distributions in the combined Run~2 fit. 
The shaded band represents the post-fit PDF uncertainty. 
Right: Values of $\sineff$ measured in each of the four channels using the full Run~2 data sample (upper)
and for different PDF sets (lower). }\label{fig:ak6}
\end{figure}

Next, we fit the measured $A_4(Y, M)$ with the $\sineff$  and PDF templates. 
The comparison of these measurements with the corresponding best-fit distributions is shown in Figure~\ref{fig:ak6}~(left), 
which illustrates an excellent fit quality, with $\chi^2_\mathrm{min}/\mathrm{ndf} = 61.3/62$. 
As a cross-check, we also extract the $\sineff$ by directly fitting the \cs distributions with the floating 
$\sineff$ and PDF templates. 
The best-fit $\sineff$ values extracted with these three methods for different channels and PDF sets are displayed 
in Figure~\ref{fig:ak6}~(right). The results agree well within the corresponding uncertainties. 
The figure also shows that profiling the PDFs significantly reduces both the individual PDF uncertainties and the spread 
of the central $\sineff$ values extracted with different PDFs. 

\begin{figure}[!htb]
\centering
\includegraphics[width=0.50\linewidth]{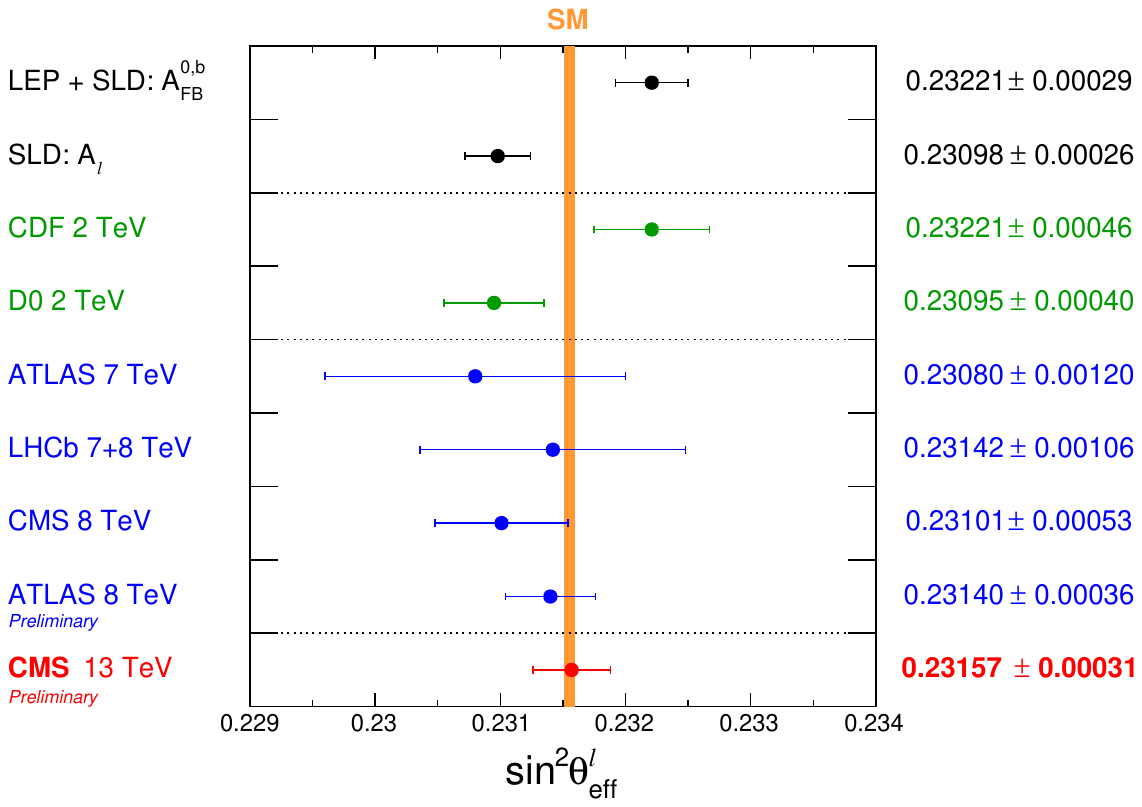}
\caption[]{Comparison of the $\sineff$ values measured in this analysis
with previous measurements and the SM prediction.}
\label{fig:ak8}
\end{figure}

Figure~\ref{fig:ak8} shows the comparison of the measured $\sineff$ with the previous results and the SM prediction. 
The total uncertainty of this measurement is the smallest among the hadron-collider measurements. 
The measurement agrees well with the SM prediction and the central value is close 
to the average of the two most precise LEP and SLD results. 

\section{Summary}

A precise measurement of $\sineff$ was performed using Run~2 
data samples of pp collisions at $\sqrt{s} = 13$\,TeV.
Using the CT18Z set of PDFs, which covers best the central values extracted with the other global PDFs, 
we have obtained
$$\sineff = 0.23157 \pm 0.00031,$$
with the total uncertainty dominated by PDFs.
The measured $\sineff$ value is in good agreement with the SM prediction, 
and is the most precise among the hadron-collider measurements.
The precision is also comparable to that of the two most precise measurements
performed in electron-positron collisions at LEP and SLD, 
with respective uncertainties of 0.00026 and 0.00029.
We have also measured the $A_4$  coefficient as a function of the dilepton mass and rapidity,
which can be used in combination with other LHC measurements
and in improvements of the $\sineff$ measurement with future PDF sets.

\section*{Acknowledgements}
The author wishes to thank the organizers of the Moriond conference. This work was supported in part by
the US Department of Energy award DE-SC0008475.

\end{document}